\documentclass[english,reprint]{revtex4-2}
\usepackage[T1]{fontenc}
\usepackage[latin9]{inputenc}
\usepackage{babel}
\usepackage{amsmath}
\usepackage{amssymb}
\usepackage{graphicx}
\usepackage[unicode=true]
{hyperref}

\makeatletter
\hypersetup{colorlinks}

\makeatother

\begin{document}
    \title{Supersolid Gap Soliton in a Bose-Einstein Condensate and Optical Ring
        Cavity coupling system}
    \author{Jieli Qin}
    \email{qinjieli@126.com; 104531@gzhu.edu.cn}
    
    \address{School of Physics and Materials Science, Guangzhou University, 230
        Wai Huan Xi Road, Guangzhou Higher Education Mega Center, Guangzhou
        510006, China}
    \author{Lu Zhou}
    \email{lzhou@phy.ecnu.edu.cn}
    
    \address{Department of Physics, School of Physics and Electronic Science, East
        China Normal University, Shanghai 200241, China}
    \address{Collaborative Innovation Center of Extreme Optics, Shanxi University,
        Taiyuan, Shanxi 030006, China}
    \begin{abstract}
        The system of a transversely pumped Bose-Einstein condensate (BEC) coupled to a lossy ring cavity can favor a supersolid steady state. Here we find the existence of supersolid gap soliton in such a driven-dissipative system. By numerically solving the mean-field atom-cavity field coupling equations, gap solitons of a few different families have been identified. Their dynamical properties, including stability, propagation and soliton collision, are also studied. Due to the feedback atom-intracavity field interaction, these supersolid gap solitons show numerous new features compared with the usual BEC gap solitons in static optical lattices.
    \end{abstract}
    \maketitle
    
    \section{Introduction\label{sec:Introduction}}
    
    Due to the effect of dispersion, a wave packet would suffer a spatially
    spreading during its time evolution. However, when there also exists
    appropriate nonlinearity in the system, the dispersion spreading can
    be suppressed, and give rise to a non-spreading localized wave packet---soliton
    \cite{Drazin1989Solitons}. In a Bose-Einstein condensate (BEC) system,
    the nonlinearity due to attractive inter-atom contact interaction
    can well balance the dispersion spreading, and support a matter wave
    soliton, which is particularly called as bright soliton \cite{Ruprecht1995Time,PerezGarcia1998Bose,Khaykovich2002Formation,Strecker2002Formation,Strecker2003Bright,Cornish2006Formation,Billam2013Bright}.
    While for the repulsive interaction, instead of a non-spreading wavepacket,
    it only supports a localized atomic density dip (i.e., absence of
    atoms) on the background BEC density profile, which is usually named
    as dark soliton \cite{Jackson1998Solitary,Dum1998Greation,Burger1999Dark,Denschlag2000Generating,Anderson2001Watching,Wu2002Controlled,Frantzeskakis2010Dark}
    (these solitons are originally studied in the field of nonlinear optics,
    so according to the brightness of the light pulses, they are described
    by the words ``bright'' and ``dark'' \cite{Malomed2005Spatiotemporal,Chen2012Optical,Song2019Recent}).
    When BEC is loaded into a periodical optical lattice potential, its
    dispersion property can be substantially changed, as a result, even
    for the repulsive interaction there can also exist a bright soliton.
    Since the chemical potential of such a soliton falls into the energy
    gaps of the optical lattice potential, it is given the name of gap
    soliton \cite{Alfimov2002Wannier,Louis2003Bose,Efremidis2003Lattice,Eiermann2004Bright,Dabrowska2004Interaction,Anker2005Nonlinear,Sakaguchi2005Matter,Mayteevarunyoo2006Stability,Morsch2006Dynamics,Sakaguchi2006Gap,Richter2007Long,Wang2008Localized,Sivan2008Qualitative,Zhang2009Composition,Zhang2009Gap,Alexander2005Soliton,Alexander2006Self,Sakaguchi2010Solitons,Kartashov2011Solitons,Dror2013Stability,Su2021Creating}.
    In passing, we also mention that although gap solitons usually refer
    to the bright ones, there do exist dark gap solitons \cite{Kivshar1994Dark,Cheng2018Dark,Zeng2019Gap,Li2021Dark}
    which are not the concerns of this paper.
    
    Supersolid is an unusual state of matter that simultaneously behaves
    as both a crystalline solid and a superfluid \cite{Prokofev2007What,Balibar2010The,Kuklov2011How,Boninsegni2012Supersolid,Chan2013Overview}.
    Originally, it was predicted for helium as early as the middle of
    the 20th century \cite{Gross1957Unified,Thouless1969The,Andreev1969Quantum,Chester1970Speculations,Leggett1970Can},
    but until now it still has not been observed undoubtedly \cite{Kim2004Observation,Kim2004Probable,Day2007Low,Kim2012Absence,Maris2012Effect}.
    In recent years, the highly controllable atomic quantum gas brings
    new vitality to the supersolid studies. It has been experimentally
    realized in several different types of systems. The ground state of a
    spin-orbit coupled BEC can fall into a stripe phase exhibiting supersolid
    properties \cite{Li2017A,Bersano2019Experimental,Putra2020Sptital}.
    In dipolar BEC, the balance between long range dipole-dipole interaction
    and short range contact interaction gives rise to the emergence of
    arrays of quantum droplets, i.e. dipolar supersolid \cite{Tanzi2019Observation,Tanzi2019Supersolid,Guo2019The,Bottcher2019Transient,Chomaz2019Long,Norcia2021Tow,Sohmen2021Birth}.
    The cavity-mediated interaction also can lead to BEC supersolid. Two
    different schemes have already been experimentally reported, one of
    them couples the BEC to two crossed linear optical cavities \cite{Leonard2017Supersolid,Leonard2017Monitoring},
    while the other one uses a ring cavity \cite{Mivehvar2018Driven,Schuster2020Supersolid}.
    
    In this work, we are interested in the driven-dissipative supersolid
    BEC realized by the ring cavity scheme, for details of the physical
    model see section \ref{sec:Model} or the original paper \cite{Mivehvar2018Driven}.
    In this scheme, the BEC collectively scatters the pumping photons
    into the cavity, and results in a superradiant optical lattice,
    which backwardly drives the BEC into a supersolid state. As having
    been pointed out in the first paragraph of this section, the simultaneous
    existence of an optical lattice and interaction nonlinearity
    would lead to gap solitons in BEC. In this vein of thought, we propose that there would
    exist supersolid gap solitons in a BEC and ring cavity
    coupling system. The main objectives of this work are finding out such
    soliton solutions, and studying their basic properties.
    
    The rest contents of this paper are organized as follows: In section
    \ref{sec:Model}, we briefly describe the considering system, and
    present the theoretical formulas to handle its steady state and dynamical
    evolution. The next section \ref{sec:Results} shows the main results
    of this paper. It is split into four subsections, which deal with
    the gap soliton solutions (\ref{subsec:GapSolitonSolutions}), their
    stability (\ref{subsec:Stability}), mobility (\ref{subsec:Mobility})
    and collision (\ref{subsec:Collision}) properties respectively. At
    last, we summarize the paper in section \ref{sec:Summary}.
    
    \section{Model\label{sec:Model}}
    \begin{figure}
        \begin{centering}
            \includegraphics{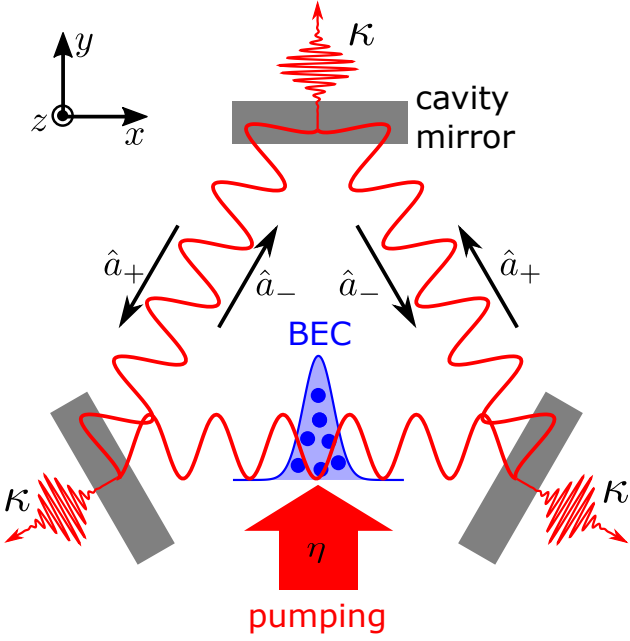}
            \par\end{centering}
        \caption{Schematic diagram of the considering atomic BEC and optical ring cavity
            coupling system. A quasi-one-dimensional BEC is loaded in an optical
            ring cavity along its axis. By transversely pumping the BEC using
            a standing wave laser (pumping strength $\eta$), light fields are
            built up in the cavity due to superradiant scattering of pumping photons
            into the two counterpropagating ring cavity optical modes ($\hat{a}_{\pm}e^{\pm ik_{c}x}$),
            and reach a steady state on account of the cavity loss (loss rate
            $\kappa$). Interference between lights of the two cavity modes forms
            an optical lattice. Backwardly, this optical lattice drives the BEC
            to a supersolid gap soliton state \cite{Mivehvar2018Driven}.\label{fig:Diagram}}
    \end{figure}
    
    As schematically shown in figure \ref{fig:Diagram}, in this work
    we consider an atomic BEC and optical ring cavity coupling system.
    The BEC is loaded in the ring cavity along the cavity axis, and is
    tightly trapped in the transverse ($y,z$) direction, such that only
    the longitudinal ($x$) direction dynamics need to be considered.
    When transversely illuminating a standing-wave laser on the BEC atoms
    (Rabi frequency $\Omega_{0}$ and detuning $\Delta_{a}$), light fields
    are built up in the cavity due to the scattering of pumping photons
    into the two counterpropagating ring cavity optical modes ($\hat{a}_{\pm}e^{\pm ik_{c}x}$
    with $\hat{a}_{\pm}$ being the annihilation operators and $k_{c}$being
    the wavenumber). Backwardly, the induced cavity light fields will
    also interact with the BEC atoms (strength ${\cal G}_{0}$). Such
    a system can be described by Hamiltonian \cite{Mivehvar2018Driven}
    \begin{align}
        \hat{H} & =-\hbar\Delta_{c}\left(\hat{a}_{+}^{\dagger}\hat{a}_{+}+\hat{a}_{-}^{\dagger}\hat{a}_{-}\right)+\int\hat{\psi}^{\dagger}\left(x\right)H_{a}\hat{\psi}\left(x\right)dx\nonumber \\
        & +\frac{1}{2}\int\hat{\psi}^{\dagger}\left(x\right)\hat{\psi}^{\dagger}\left(x'\right)V\left(x-x'\right)\hat{\psi}\left(x'\right)\hat{\psi}\left(x\right)dxdx',\label{eq:Hamiltonian_eff}
    \end{align}
    where $\hbar$ is the Planck constant, $\Delta_{c}$ is the detuning
    between the cavity modes and pump laser, $\hat{\psi}$ is the field
    operator of the BEC. The first term describes the two counter-propagating
    cavity modes. The last term describes the interaction between BEC
    atoms. Considering the one-dimensional effective contact interaction
    \cite{Salasnich2002Effective}, it is $V\left(x-x'\right)=g\delta\left(x-x'\right)$,
    where $g=2\hbar\omega_{\bot}a_{s}$ describes the interaction strength,
    with $a_{s}$ being the s-wave scattering length, $\omega_{\bot}$
    referring to the frequency of transverse confinement harmonic potential.
    When $g<0$, it represents an attractive interaction, while for $g>0$,
    it is a repulsive interaction. In this work, we consider the case
    of repulsive interaction. The middle term accounts for the kinetic
    and optical potential energy of the BEC, $H_{a}$ is the corresponding
    single particle Hamiltonian
    
    \begin{align}
        H_{a} & =\frac{\hat{p}_{x}^{2}}{2m}+\hat{V}_{ac}+\hat{V}_{ap},\label{eq:Hamiltonian_atom}
    \end{align}
    with
    \begin{equation}
        \hat{V}_{ac}=\hbar U\!\left[\hat{a}_{+}^{\dagger}\hat{a}_{+}+\hat{a}_{-}^{\dagger}\hat{a}_{-}+\left(\hat{a}_{+}^{\dagger}\hat{a}_{-}e^{-2ik_{c}x}+\mathrm{h.c.}\right)\right],\label{eq:V_ac}
    \end{equation}
    \begin{equation}
        \hat{V}_{ap}=\hbar\eta\left(\hat{a}_{+}e^{ik_{c}x}+\hat{a}_{-}e^{-ik_{c}x}+\mathrm{h.c.}\right).\label{eq:V_ap}
    \end{equation}
    Here, $m$ is the atomic mass, $\hat{p}_{x}=-i\hbar\partial_{x}$
    is the momentum operator along the $x$ direction, hence $\hat{p}_{x}^{2}/(2m)$
    is the kinetic energy term. The optical potential can be split into
    two parts. The part caused by two-photon scattering between the two
    cavity modes is denoted as $\hat{V}_{ac}$, its strength is $\hbar U=\hbar\mathcal{G}_{0}/\Delta_{a}$.
    The other part $\hat{V}_{ap}$ is caused by two-photon scattering
    between pump and cavity modes, and its strength is $\hbar\eta=\hbar\mathcal{G}_{0}\Omega_{0}/\Delta_{a}$.
    In other words, this term describes the pumping of the system, so
    $\eta$ would also be called an effective pumping strength. In the
    following contents, we will use the natural unit $m=\hbar=k_{c}=1$ for
    simplicity of formulae.
    
    Within the mean-field theory \cite{Mivehvar2021Cavity}, the quantum
    operators are approximated by their mean values. The dynamical equations
    governing these mean-field variables can be obtained by taking mean
    values of the corresponding Heisenberg equations, and they are
    \begin{equation}
        i\frac{\partial}{\partial t}\alpha_{\pm}=\left(-\Delta_{c}+UN-i\kappa\right)\alpha_{\pm}+UN_{\pm2}\alpha_{\mp}+\eta N_{\pm1},\label{eq:Meanfield_cavity}
    \end{equation}
    \begin{equation}
        i\frac{\partial}{\partial t}\psi=\left[-\frac{1}{2}\frac{\partial^{2}}{\partial x^{2}}+\mathcal{V}_{\mathrm{eff}}\left(x\right)\right]\psi+\left|\psi\right|^{2}\psi,\label{eq:Meanfield_atom}
    \end{equation}
    with
    \begin{equation}
        N=\int\left|\psi\left(x\right)\right|^{2}dx,\label{eq:AtomNumber}
    \end{equation}
    \begin{equation}
        N_{\pm1}=\int\left|\psi\left(x\right)\right|^{2}e^{\mp ix}dx,\label{eq:N1}
    \end{equation}
    \begin{equation}
        N_{\pm2}=\int\left|\psi\left(x\right)\right|^{2}e^{\mp2ix}dx,\label{eq:N2}
    \end{equation}
    \begin{equation}
        \mathcal{V}_{\mathrm{eff}}\left(x\right)=\mathcal{V}_{ac}\left(x\right)+\mathcal{V}_{ap}\left(x\right),\label{eq:Veff}
    \end{equation}
    \begin{equation}
        \mathcal{V}_{ac}=U\left(\left|\alpha_{+}\right|^{2}+\left|\alpha_{-}\right|^{2}\right)+U\left(\alpha_{+}^{*}\alpha_{-}e^{-2ix}+\mathrm{c.c.}\right),\label{eq:VacMean}
    \end{equation}
    \begin{equation}
        \mathcal{V}_{ap}=\eta\left(\alpha_{+}e^{ix}+\alpha_{-}e^{-ix}+\mathrm{c.c.}\right),\label{eq:VapMean}
    \end{equation}
    being some auxiliary quantities to make the dynamical equations
    compact. Here, we have introduced the cavity loss with rate $\kappa$
    phenomenologically. And, we also have scaled the BEC wave function
    with the inter-atom interaction character length $l=\omega_{\bot}a_{s}m/\left(2\hbar k_{c}^{2}\right)$,
    $\psi\rightarrow\text{\ensuremath{\psi}}/\sqrt{l}$, therefore in
    equation (\ref{eq:Meanfield_atom}) the coefficient before the nonlinear
    interaction term is simplified to 1. Under such a scaling, the normalization
    constant $N$ should be interpreted as a scaled atom number. However,
    without leading to any misunderstanding, literally we will still call
    it ``atom number'' for convenience in the following contents.
    
    Due to the balance between the pumping and cavity loss, the system
    will reach a steady state which can be mathematically obtained by
    letting $\partial_{t}\alpha_{\pm}=0$, and $\psi\left(x,t\right)=\psi\left(x\right)e^{-i\mu t}$
    with $\mu$ being the chemical potential. Inserting them into equations
    (\ref{eq:Meanfield_cavity}) and (\ref{eq:Meanfield_atom}), one gets
    the following time-independent equations for steady state
    
    \begin{equation}
        \mu\psi\left(x\right)=\left[-\frac{1}{2}\frac{\partial^{2}}{\partial x^{2}}+\mathcal{V}_{\mathrm{eff}}\left(x\right)\right]\psi\left(x\right)+\left|\psi\left(x\right)\right|^{2}\psi\left(x\right),\label{eq:steadypsi}
    \end{equation}
    \begin{equation}
        \alpha_{+}=-\frac{\left(-\Delta_{c}+UN-i\kappa\right)\eta N_{+1}-\eta UN_{+2}N_{-1}}{\left(-\Delta_{c}+UN-i\kappa\right)^{2}-U^{2}N_{-2}N_{+2}},\label{eq:alpha_p}
    \end{equation}
    
    \begin{align}
        \alpha_{-} & =-\frac{\left(-\Delta_{c}+UN-i\kappa\right)\eta N_{-1}-\eta UN_{-2}N_{+1}}{\left(-\Delta_{c}+UN-i\kappa\right)^{2}-U^{2}N_{-2}N_{+2}}.\label{eq:alpha_m}
    \end{align}
    
\begin{figure}
        \begin{centering}
            \includegraphics{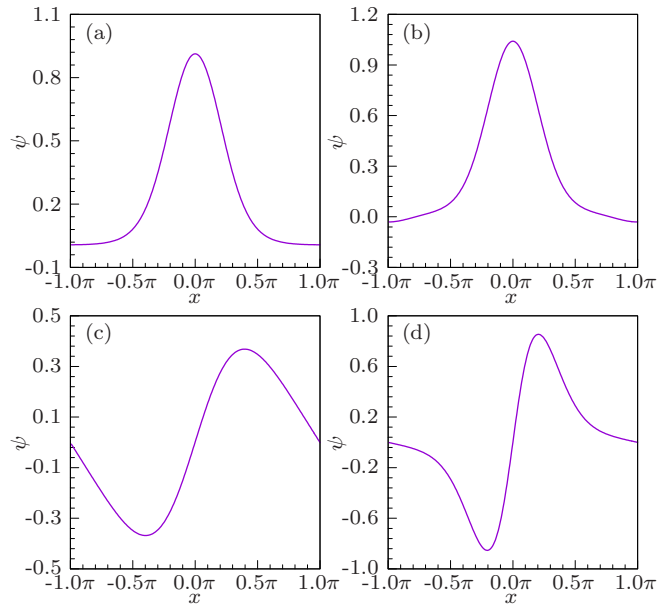}
        \end{centering}
        \caption{
            Examples of spatially periodical supersolid states. Only one spatial period is considered here. Chemical potentials corresponding to these states are respectively $\mu=-3.10$ (a), $-3.04$ (b), $0.55$ (c), and $-1.31$ (d), i.e., the same as those in panels (b,f) of figure \ref{fig:FundamentalSolitons}, and panels (b,e) of figure \ref{fig:SubfundamentalSoliton}. Parameters used are $\Delta_{c}=-1$, $U=-0.5$, $\kappa=10$, $\eta=15$ (this set of parameters will be used all along this paper). \label{fig:PeriodicalSolutions}
        }
    \end{figure}

    We see that the BEC feels an optical lattice potential from the light
    fields. Since we are considering a running wave ring cavity, the location
    of this optical lattice is not predetermined by the cavity mirrors,
    and spontaneously breaks the continuous translation symmetry.
    This optical lattice will further modulate the BEC atomic density, that is, it will
    drive the BEC to a spatially periodical supersolid state \cite{Mivehvar2018Driven,Schuster2020Supersolid}.
    For some examples of such states, one can see figure \ref{fig:PeriodicalSolutions}.
    
    It is well known that when BEC with repulsive contact interaction
    is loaded in an optical lattice, even though the system is not well
    bounded, a kind of localized wavepacket, i.e., gap soliton, can also
    exist due to the balance between repulsive interaction and anomalous
    dispersion \cite{Eiermann2004Bright}. Here, a supersolid optical
    lattice is also built up, thus we guess that gap soliton would also
    exist in the now considering supersolid system.  Next, we try to
    find such supersolid gap soliton solutions, and examine their stability,
    mobility, and collision properties.
    
    \section{Results\label{sec:Results}}
    \subsection{Gap Soliton Solutions\label{subsec:GapSolitonSolutions}}
    \begin{figure*}
        \begin{centering}
            \includegraphics{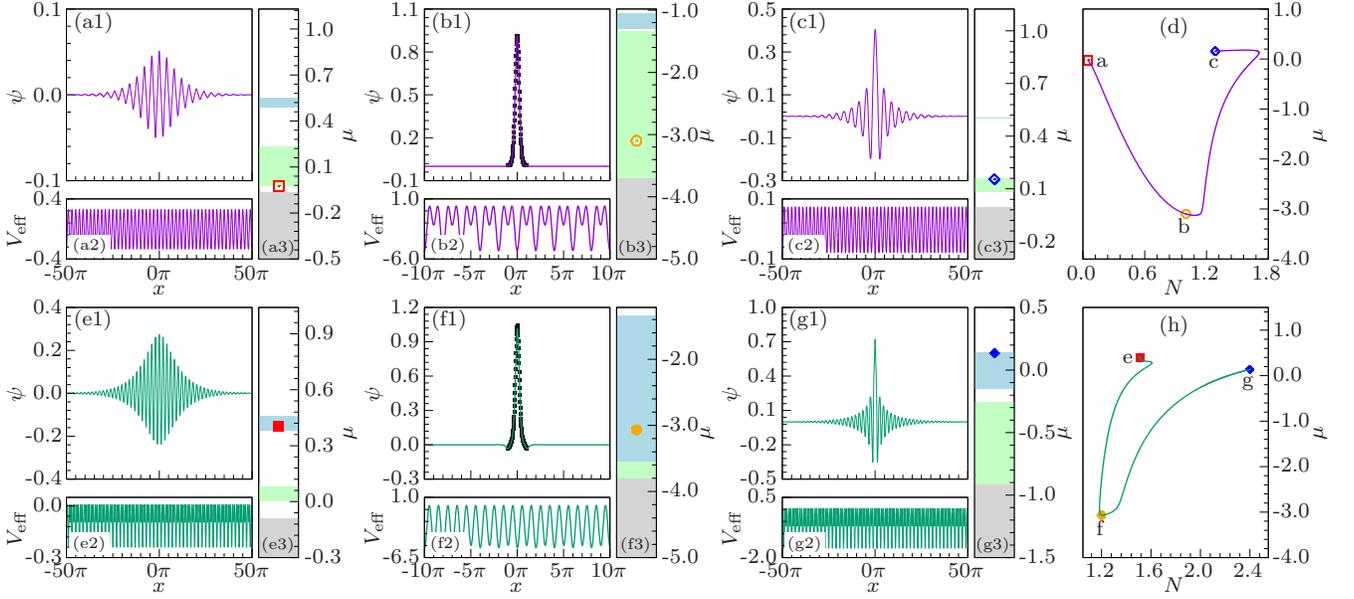}
            \par\end{centering}
        \caption{Fundamental gap solitons in the first (a-d) and second (e-h) energy
            gaps. Panels (a1-c1) and (e1-g1): wave functions $\psi\left(x\right)$
            of the fundamental gap solitons.
            In panels (b1,f1), the black points represent the periodical supersolid states [which have previously been shown in panels (a,b) of figure \ref{fig:PeriodicalSolutions}] with the same chemical potential.
            Panels (a2-c2) and (e2-g2): corresponding
            effective optical lattice potentials $V_{\mathrm{eff}}\left(x\right)$.
            Panels (a3-c3) and (e3-g3): corresponding energy band structures.
            The gray filled region is the semi-infinite energy gap, the light-green filled region is the first energy gap, the light-blue filled region is the second energy gap, while the unfilled regions
            are the energy bands (some of the energy bands and gaps are too narrow
            to be seen very clearly). The colored points mark the chemical potential
            $\mu$ of the corresponding solutions. In panels (a,e), the chemical potentials lie close to the lower edge of the gap; in panels (b,f), they lie deep in the gap; and in panels (c,g), they lie near
            the upper edge of the gap. Panels (d) and (h): relation between chemical
            potential $\mu$ and atom number $N$ for fundamental gap soliton
            in the first (d) and second (h) energy gaps. The values of ($N,\mu$)
            corresponding to solutions of panels (a-c) and (e-g) are marked on
            the curve with the colored points which have point style the same
            as those in panels (a3-c3) and (e3-g3), and at the same time also
            explicitly labeled. \label{fig:FundamentalSolitons}}
    \end{figure*}
    
    \begin{figure*}
        \begin{centering}
            \includegraphics{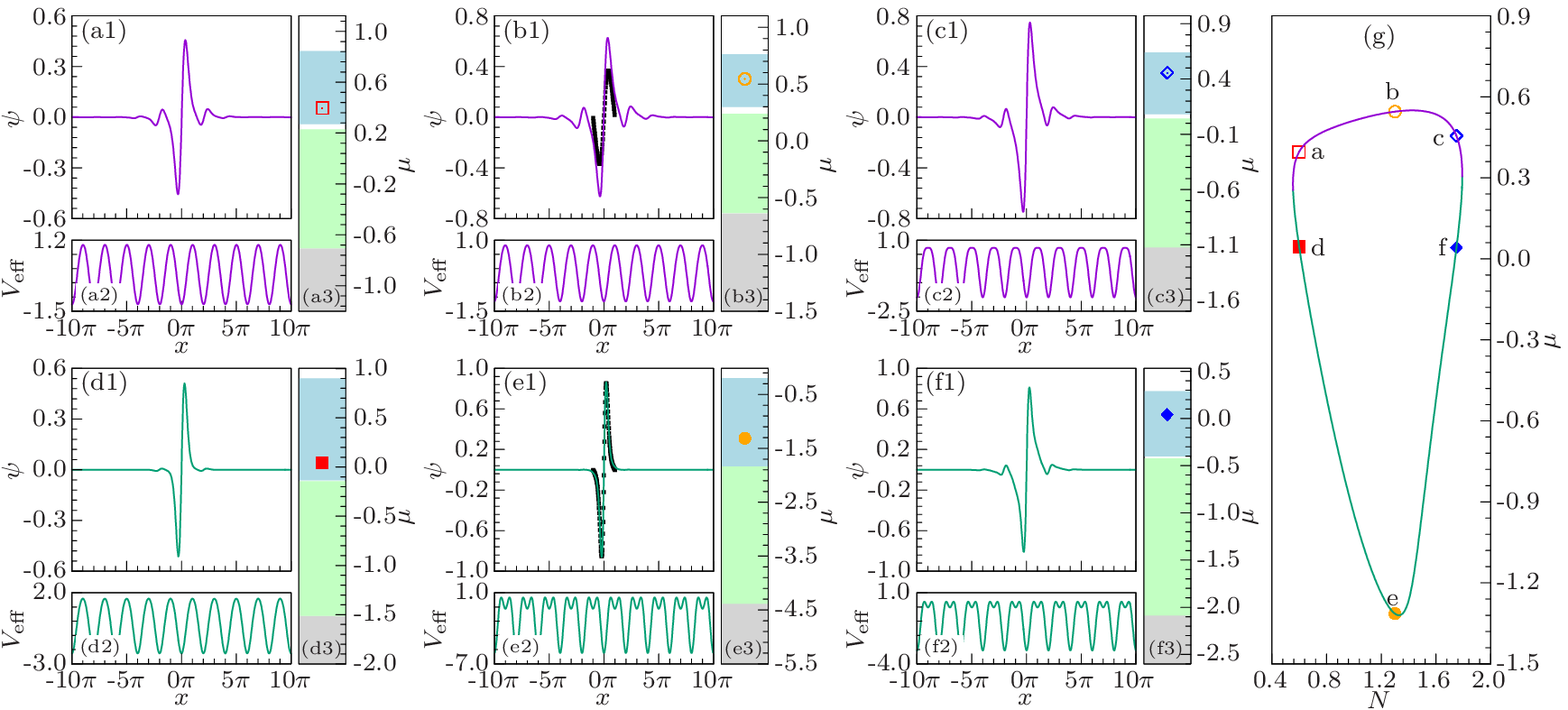}
            \par\end{centering}
        \caption{Sub-fundamental gap solitons in the second energy gap. Panels (a1-f1),
            (b2-f2), (a3-f3) and (g) are wavefunctions, effective optical lattice,
            energy band structure and $N$-$\mu$ curve respectively, and they
            are plotted the same way as figure \ref{fig:FundamentalSolitons}.
            In panels (b1,e1), the black points represent the periodical supersolid states [which have previously been shown in panels (c,d) of figure \ref{fig:PeriodicalSolutions}] with the same chemical potential.
            The $N$-$\mu$ curve of this family of soliton is a closed loop,
            the upper and lower part is plotted with violet and green color respectively.
            The solutions of panels (a-c) have values of ($N,\mu$) on the upper
            part of the $N$-$\mu$ loop, while the solutions of panels (d-f)
            have values of ($N,\mu$) on the lower part, as labeled in panel (g).
            \label{fig:SubfundamentalSoliton}}
    \end{figure*}
    
    We find gap soliton by numerically solving the discretized version (the derivative is approximated by second-order central difference) of equations
    (\ref{eq:steadypsi}, \ref{eq:alpha_p}, \ref{eq:alpha_m}), starting from an initial guessing wave function $\psi_{0}=\sum_{i=1,I}A_{i}\mathrm{sech}\left[\left(x-x_{i}\right)/\sigma_{i}\right]$
    , where $A_{i}$, $x_{i}$, $\sigma_{i}$ and $I$ are numerically tunable
    parameters. $I$ determines the number of sub-wavepackets (peaks) of the soliton, $x_i, \sigma_i, A_i$ are the location, width and amplitude of each sub-wavepacket. We expect that each sub-wavepacket will fit in a lattice site, thus $\sigma_i$ is typically set to a value several times smaller than the spatial period of the optical lattice (since the dimension-less optical lattice period is $2\pi$, we typically set $\sigma_i$ in the range of $0.3 \sim 2$). Parameter $A_i$ is typically set to a value in the range of $0.1 \sim 1$, such that the contact interaction is obvious, but smaller than the depth of optical lattice, and a self-bounded bright soliton solution would be possible.
    The same typical physical parameters $\Delta_{c}=-1$, $U=-0.5$, $\kappa=10$, $\eta=15$  will be used all along this paper.
    
    In figure \ref{fig:FundamentalSolitons}, we show some examples
    of fundamental soliton wavefunctions in the first [panels (a1-c1)]
    and second [panels (e1-g1)] energy gaps, together with their corresponding
    effective optical lattice potentials [panels (a2-c2; e2-g2)] and
    energy band structures [panels (a3-c3; e3-g3)]. When the soliton's
    chemical potential lies deep in the energy gap, its wavefunction is
    well localized in only one lattice site [panels (b1,f1)]. As the chemical potential
    moves towards the edge of the energy gap, oscillating-decay tails grow
    out on both sides of the central peak of the wavefunction. When the
    chemical potential becomes very close to the gap edge, the tail grows
    very heavy, see panels (a1,e1; c1,g1). Compared with the close to top
    gap edge case [panels (c1,g1)], near the bottom edge of the energy gap, we
    found that the tail decays much slower, so that the height of the
    tail peaks becomes almost as high as the central main peak [panels
    (a1,e1)].
    
    The relation between chemical potential and atom number plays an important
    role in studying gap solitons. It is the basis for classifying
    gap solitons---a distinct family of gap solitons is usually identified by
    a continuous $N$-$\mu$ curve \cite{Efremidis2003Lattice,Mayteevarunyoo2006Stability}.
    For the usual gap solitons in a static optical lattice, this relation is very
    simple. As the atom number increases, the repulsive interaction leads
    the chemical potential to grow gradually, i.e., $\mu$ is a monotonic
    increasing function of $N$ \cite{Efremidis2003Lattice,Mayteevarunyoo2006Stability}.
    Here, the optical lattice potential is built up by pumping the atomic
    BEC. When the quantum state of BEC changes, the optical lattice potential also changes accordingly. The potential energy will also have an affection on the chemical
    potential. Therefore, we found that the relationship between $N$ and
    $\mu$ becomes more complex, see panels (d,h) of figure \ref{fig:FundamentalSolitons}.
    For the well-localized soliton with chemical potential deep in the
    energy gap [panels (b,f)], a deep optical lattice is produced 
    (hence the corresponding energy gap is relatively wide), the potential energy
    makes the chemical potential take a large negative value. Near the
    edges of the energy gap [panels (a,e;c,g)], 
    the soliton wavefunction extends much wider.
    According to formulas (\ref{eq:N1}--\ref{eq:alpha_m}), this will lead
    to smaller values of $N_{\pm1,\pm2}$, consequently a shallower optical
    lattice (hence the corresponding energy gap becomes narrower) 
    and close to zero chemical potential. 
    Therefore, roughly speaking
    the $N$-$\mu$ curve takes a ``V'' shape, as shown in panels (d,h). In both
    these two panels, the right arms of the ``V'' shapes have a positive
    dependence of $\mu$ on $N$. This can be understood by the fact that
    as $N$ increases the repulsive interaction becomes stronger, and
    at the same time as moves towards the gap edge, the lattice potential
    also becomes shallower. These two mechanisms both lead to the positive
    dependence of $\mu$ on $N$. However, in these two panels, the left
    arms of the ``V'' shapes have opposite slopes. In panel (d), as $N$
    increases the repulsive interaction energy will increase accordingly.
    Meanwhile, the induced optical lattice becomes deeper, which
    leads to a decrease of potential energy. Since the atom number
    $N$ is small in this regime, it is the decrease of potential energy that
    plays the main role, thus in total $\mu$ has a negative dependence on $N$.
    In panel (h), even for the left arm, the atom number is also considerably
    large, so this time the repulsive interaction energy dominants the
    potential energy, and the chemical potential $\mu$ will increase
    with the atom number $N$. Lastly, we emphasize that the above discussions are
    only roughly valid, the dependence of $\mu$ on $N$ is in fact very
    complex, in detail, we also found that the $N$-$\mu$ curves may anomalously
    bend near the gap edge [the right end of panel (d), and the left
    end of panel (h)]. This anomalous bent reflects the nonlinear feature
    of the superradiant optical lattice, the similar phenomenon also happens
    for matter-wave solitons in other nonlinear optical lattices \cite{Sakaguchi2005Matter}.
    
    In the second energy gap, other than the fundamental solitons, we
    also found another family of solitons, see figure \ref{fig:SubfundamentalSoliton}.
    The wavefunction of this type of soliton is very similar to the first excited state of a (harmonically) trapped particle. It possesses the odd parity symmetry, at the lattice bottom the wavefunction takes a zero value, and there are two main peaks (one is positive valued, while the other is negative valued) around this zero point. We distinguish this family of solitons as sub-fundamental solitons because their wavefunctions have the same feature as the sub-fundamental solitons reported in Ref. \cite{Mayteevarunyoo2006Stability},
    where gap solitons in a static optical lattice are studied. 
    The ($N,\mu$) data points of this family of solitons form a closed loop,
    see panel (g). The chemical potential of this family of solitons can
    not take values very close to the gap edge, therefore their oscillating-decay
    tails are always not very heavy. For the same atom number $N$,
    the solitons with chemical potential on the upper half part of the
    $N$-$\mu$ curve [panels (a-c)] are wider, and at the same time
    have a heavier oscillating-decay tail, compared to their lower partners
    [panels (d-f)].

    \begin{figure}
        \begin{centering}
            \includegraphics{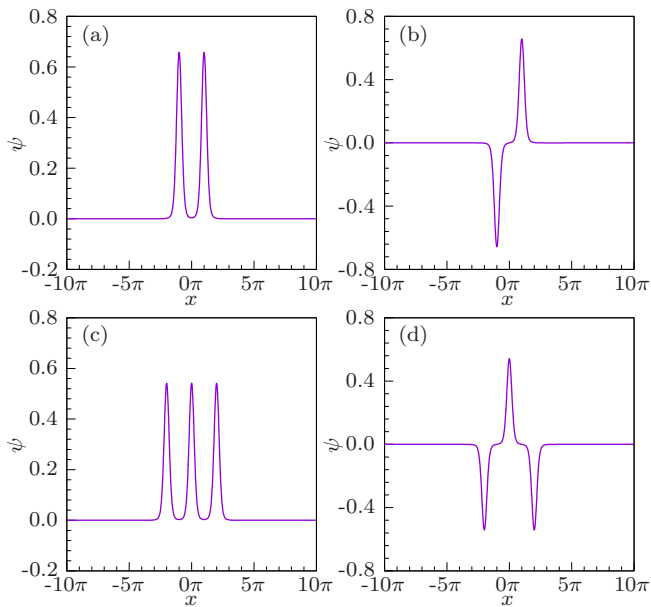}
            \par\end{centering}
        \caption{Examples of high order gap solitons in the first energy gap. They
            can be interpreted as combinations of two (a,b) or three (c,d) fundamental
            solitons with the same (a,c) or opposite (b,d) phases. \label{fig:HighOrderGap1Solitons}}
    \end{figure}
    
    \begin{figure}
        \begin{centering}
            \includegraphics{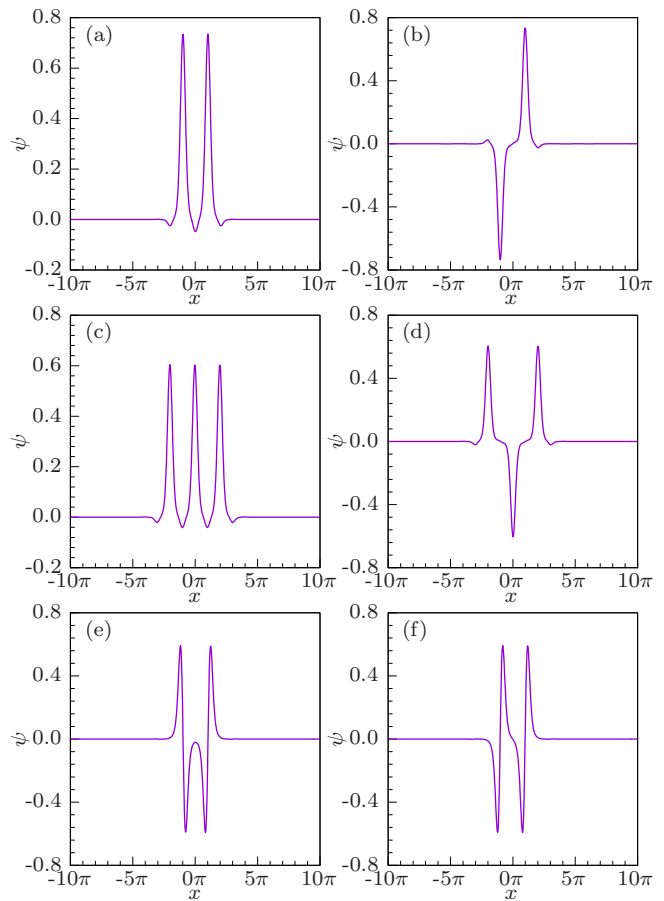}
            \par\end{centering}
        \caption{Examples of high order gap solitons in the second energy gap. They
            can be interpreted as combinations of two (a,b,e,f) or three (c,d)
            fundamental (a-d) or sub-fundamental (e,f) solitons with the same
            or opposite phases. \label{fig:HighOrderGap2Solitons}}
    \end{figure}
    
    \begin{figure*}
        \begin{centering}
            \includegraphics{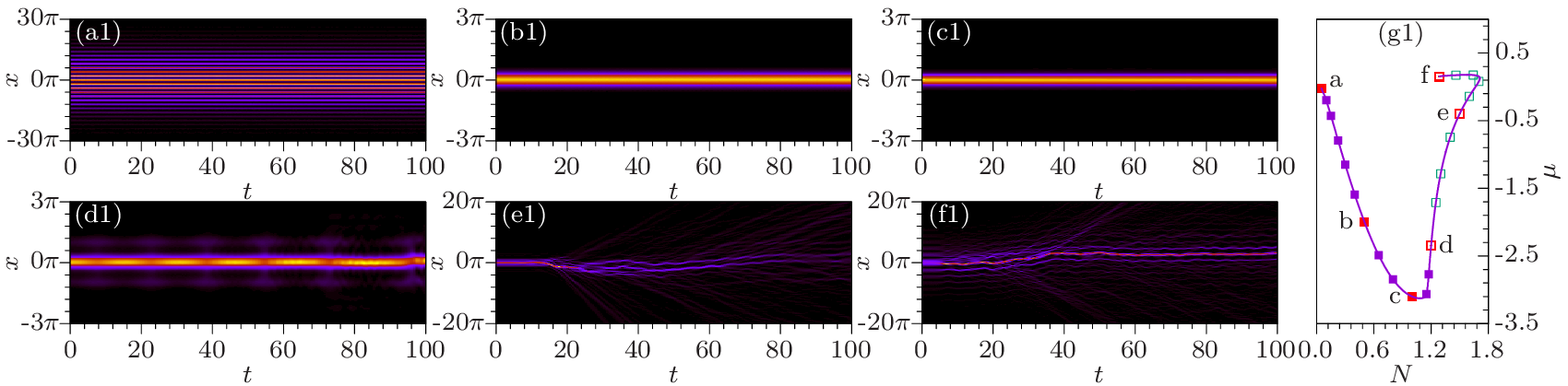}
            \par\end{centering}
        \begin{centering}
            \includegraphics{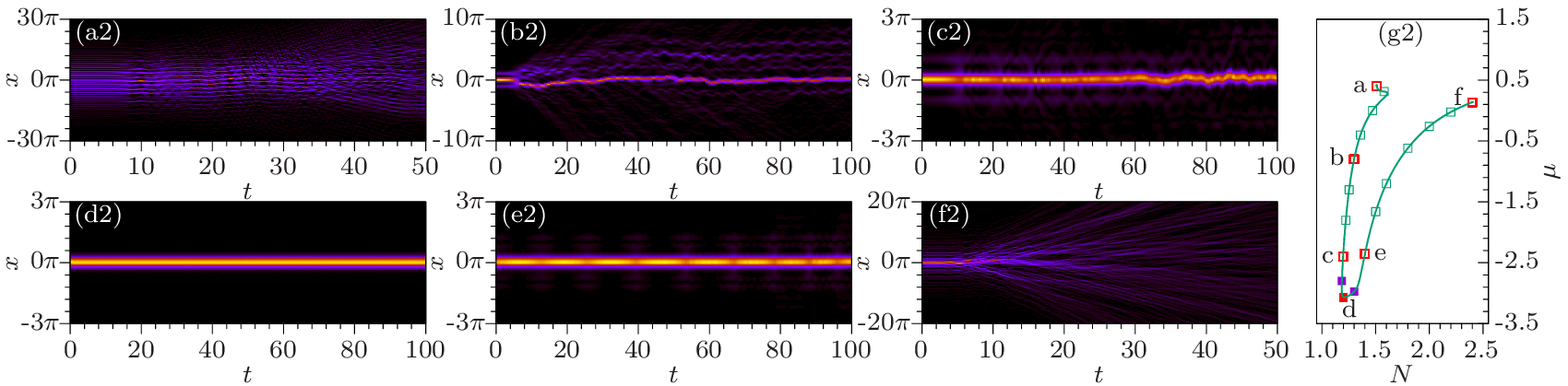}
            \par\end{centering}
        \caption{Stability of fundamental gap solitons in the first (a1-g1) and second
            (a2-g2) energy gaps. Panels (a1-f1; a2-f2): time evolutions of the atomic
            density $\left|\psi\left(x,t\right)\right|^{2}$ for some example solitons.
            Panel (g1, g2): stability of the solitons marked on the $N$-$\mu$
            curve. The solid square points represent the solitons that are stable
            during the time evolution, while the empty square points represent
            the unstable ones. The ($N,\mu$) data points corresponding to
            panels (a1-f1; a2-f2) are specially plotted with red color, and at
            the same time also have been explicitly labeled.\label{fig:StabilityFundamentalSoliton}}
    \end{figure*}
    
    \begin{figure}
        \begin{centering}
            \includegraphics{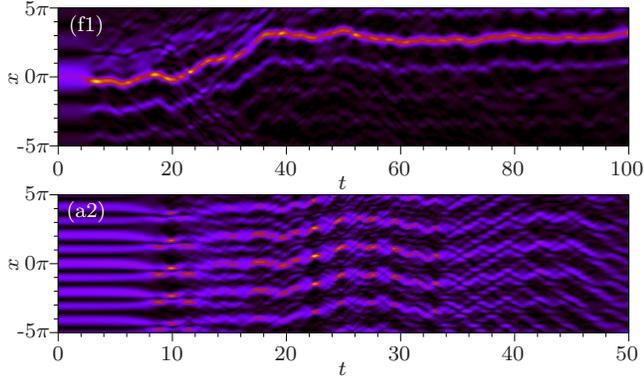}
            \par\end{centering}
        \caption{Enlargements of panels (f1) and (a2) in figure \ref{fig:StabilityFundamentalSoliton}.
            Originally, in figure \ref{fig:StabilityFundamentalSoliton}, to get
            the overall information, these two panels are plotted in a wide spatial
            range. Here, to see the details of interest, they are plotted
            within a narrow range of $x\in\left[-5\pi,5\pi\right]$. \label{fig:Enlargements}}
    \end{figure}
    
    \begin{figure*}
        \begin{centering}
            \includegraphics{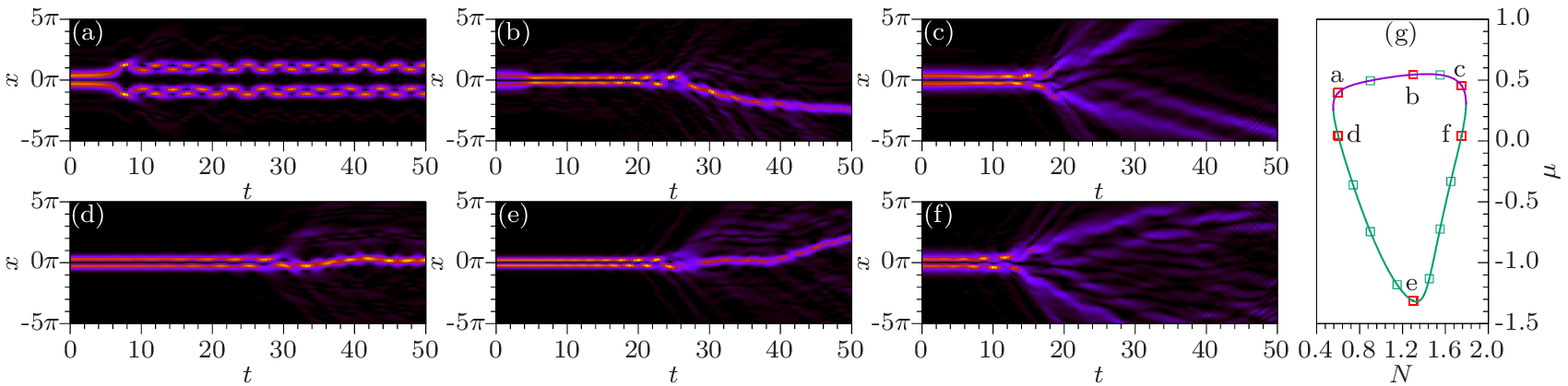}
            \par\end{centering}
        \caption{Stability of the sub-fundamental gap solitons in the second energy
            gap. The panels are plotted in a similar way as in figure \ref{fig:StabilityFundamentalSoliton}.
            All these sub-fundamental solitons are unstable. \label{fig:StabilitySubFunamental}}
    \end{figure*}
    
    There also exist many families of high order gap solitons, which
    have more than one main peaks. In figure \ref{fig:HighOrderGap1Solitons}
    (first energy gap) and figure \ref{fig:HighOrderGap2Solitons} (second
    energy gap), some examples of two, three and four peaks solutions
    are shown. Comparing them with the fundamental and sub-fundamental
    solitons, it is straightforward to conclude that these high order
    solitons can be interpreted as the superposition of fundamental
    or sub-fundamental solitons at different lattice sites.
    
    We also compared the spatially periodical supersolid states and the supersolid gap solitons. As shown in panels (b1,f1) of figure \ref{fig:FundamentalSolitons}, and panel (e1) of figure \ref{fig:SubfundamentalSoliton}, when the gap soliton is well localized in only one lattice site, it has almost the same shape as the periodical supersolid state with the same chemical potential. This indicates that the periodical supersolid state can be recognized as a chain of gap solitons. While for gap soliton with obvious tails around the main peak, its shape will evidently differ from the corresponding periodical solution, this can be seen from panel (b1) of figure \ref{fig:SubfundamentalSoliton}. Such similarities and differences between the localized gap soliton and periodical wave have also been reported in the case of static optical lattices \cite{Zhang2009Composition,Zhang2009Gap,Alexander2005Soliton,Alexander2006Self}.

    \subsection{Stability\label{subsec:Stability}}
    The stability of these gap solitons has been checked by numerically
    evolving the time-dependent equations (\ref{eq:Meanfield_cavity})
    and (\ref{eq:Meanfield_atom}), with a 5\% random perturbation being
    initially added on the soliton wavefunction $\psi_0$, i.e., $\psi\left(x,t=0\right)=\psi_0 \left[1+0.05\xi\left(x\right)\right]$, with $\xi\left(x\right)$ being random numbers uniformly distributed in the range of $\left(-1,1\right)$. The corresponding atomic density is $\left|\psi\left(x,t=0\right)\right|^2 \approx \left|\psi_0\right|^2 \left[1+0.1\xi\left(x\right)\right]$, that is the atomic density is perturbed by a magnitude of 10\%. However, the perturbation on total atom number is negligible, since the mean value is $\langle\xi\left(x\right)\rangle=0$. We don't explicitly perturb the cavity optical filed, it is dynamically determined by the BEC. 
    
    Character timescales
    of the considering system are the dispersion time and cavity loss
    time. The cavity loss time is estimated by the inverse of loss rate,
    $\tau_{1}=1/\kappa=0.1$. The dispersion time is the spreading time of
    a wavepacket due to the kinetic energy term, it is related to the
    width of the studied wavepacket. Here character width of the gap solitons
    is about one lattice length $\sigma\approx\pi$, so that the dispersion
    time is $\tau_{2}=m\sigma^{2}/\hbar\approx10$. Thus, for checking
    the stability, in the numerical simulations, we typically evolve the
    initial state to a final time $t=100$ (for the stable solitons) which
    is much longer than $\tau_{1}$ and $\tau_{2}$, or until the atomic
    density is substantially different from its initial profile (for the
    unstable solitons).
    
    The stability results for fundamental solitons in the first energy
    gap are shown in panels (a1-g1) of figure \ref{fig:StabilityFundamentalSoliton}.
    In panel (g1), the overall stability property is summarized on the
    $N$-$\mu$ curve, with the solid square points referring to stable
    solitons, while the empty squares for unstable ones. For detailed
    stability information, in panels (a1-f1), we show the time evolutions
    of the atomic density for some typical solitons. In panels
    (a1-c1), the solitons can maintain their shape during long time
    evolution. While, the other ones in panels (d1-f1) all lose their
    initial shape very quickly, however in different ways. In panel (d1),
    the breathing of atomic density between the main wave packet and
    the two tail wave packets on its two sides is excited. In panel (e1),
    the soliton suffers a severe spatial spreading during the evolution.
    In panel (f1), the soliton also suffers an overall spatially spreading,
    but less severe compared to that in panel (e1), because the atom
    density is smaller (therefore, the repulsive interaction is weaker).
    Another feature in panel (f1) is that the wide main wavepacket undergoes
    a sudden shrink at the very beginning time [this is emphasized by
    an enlarged graph in the top panel of figure \ref{fig:Enlargements}],
    then the shrunk narrow wave packet can evolve comparatively stable
    for some time. 
    
    The numerical results suggest that the stability of these gap solitons
    roughly obeys a Vakhitov-Kolokolov (VK) criterion (a negative slope
    of the $N$-$\mu$ curve, $d\mu/dN<0$) \cite{Sivan2008Qualitative,Sakaguchi2010Solitons,Kartashov2011Solitons,Dror2013Stability}.
    Here, we say ``roughly'' because of two reasons. Firstly, for the
    points already on the right-half of the $N$-$\mu$ curve, but still
    very close to the bottom of the curve, although the VK criterion is invalid, 
    the solitons also can evolve
    stably for quite a long time. This would result from that the density
    profile of these solitons changes very slightly during the time evolution,
    so that the numerical simulation fails to distinguish. 
    Secondly, at the close to gap edge anomalous bent, some
    points do have negative
    slopes, however the solitons are numerically checked to be unstable.
    This indicates that the VK criterion can not capture the unstable
    mechanism shown in panel (f1). We also would like to point out that
    for the normal repulsive interaction supported BEC gap solitons in
    static linear optical lattices, their stability obeys the anti-VK
    criterion by contrast \cite{Sivan2008Qualitative,Sakaguchi2010Solitons,Kartashov2011Solitons,Dror2013Stability}.
    This again makes a definite difference between the supersolid gap
    solitons discussed here and the normal gap solitons in static optical lattices.
    
    The stability results of fundamental solitons in the second energy gap are
    shown in panels (a2-g2) of figure \ref{fig:StabilityFundamentalSoliton}.
    In this case, the VK criterion ($d\mu/dN<0$) is fulfilled in two
    places---the close to gap edge anomalous bent, and a very narrow range
    left to the bottom of $N$-$\mu$ curve, see panel (g2). Numerically,
    we found that for the anomalous bent, the VK criterion again failed
    to predict the right stability property, i.e., the solitons are unstable during
    the time evolution, for an example, see panel (a2). And around the
    bottom of $N$-$\mu$ curve, agree with the VK criterion (also roughly,
    as having been discussed in the previous paragraph), the solitons are
    checked to be stable, see panel (d2). All the other points on the
    $N$-$\mu$ curve are checked to be referring to unstable solitons.
    On different parts of the $N$-$\mu$ curve, the unstable mechanisms
    are different. Close to the stable region, the solitons are unstable
    because of the atomic density breathing, see panels (c2,e2). Far away
    from the stable region, they suffer a spatial spreading, and the larger
    atom number leads to the severer spreading, this is can be seen
    from panels (b2, f2). These two unstable mechanisms are similar to that in
    the first energy gap case. At last, at the anomalous bent, we also
    find a new unstable mechanism. As shown in panel (a2) and its enlargement
    in the bottom panel of figure \ref{fig:Enlargements}, this soliton
    has a very heavy oscillating-decay tail, i.e., the density profile contains
    many small sub-wavepackets, it is unstable due to the interaction
    between two neighboring sub-wavepackets.
    
    All the sub-fundamental gap solitons are numerically found to be unstable,
    see figure \ref{fig:StabilitySubFunamental}. Several different unstable
    mechanisms have been found. Firstly, the two main peaks of a sub-fundamental
    soliton can merge into a single peak with a loss of the atoms, see
    panels (d,e). Secondly, for large atom number 
    sub-fundamental solitons, the strong repulsive interaction
    can lead to spatial spreading instability, see panels (c,f). Thirdly,
    in panel (b), for the soliton on the very top part of $N$-$\mu$ curve,
    at first it shrinks to another sub-fundamental soliton with two narrower
    density peaks (similar to its partner on the lower part of the $N$-$\mu$
    curve), then these two narrower peaks merge into a single one again. 
    Lastly, in panel (a), we found that at the beginning time 
    the two main peaks of the sub-fundamental soliton move away from each
    other, and then spatial oscillations of the wavepackets are excited. 
    
    Since the high order solitons can be seen as superposition
    of fundamental or sub-fundamental solitons at different lattice sites, we found that they usually have a similar stability feature as their fundamental or sub-fundamental components, i.e., when the composing solitons are stable, the high order soliton is also stable and vice versa. For example, solitons (a,b,c,d) in figures \ref{fig:HighOrderGap1Solitons} and \ref{fig:HighOrderGap2Solitons} are found to be stable, while solitons (e,f) in figure \ref{fig:HighOrderGap2Solitons} are unstable.
    
    For comparison, we also examined the stability of normal gap solitons in static periodic potentials whose amplitude and periodicity are the same as the dynamically created optical lattice in ring cavity. We found that in such a static lattice solitons can undergo a stable evolution until the final time of the numerical simulation. So, we think that the instability of the supersolid gap solitons found here is caused by the dynamical property of the optical field.
    \begin{figure}
        \begin{centering}
            \includegraphics{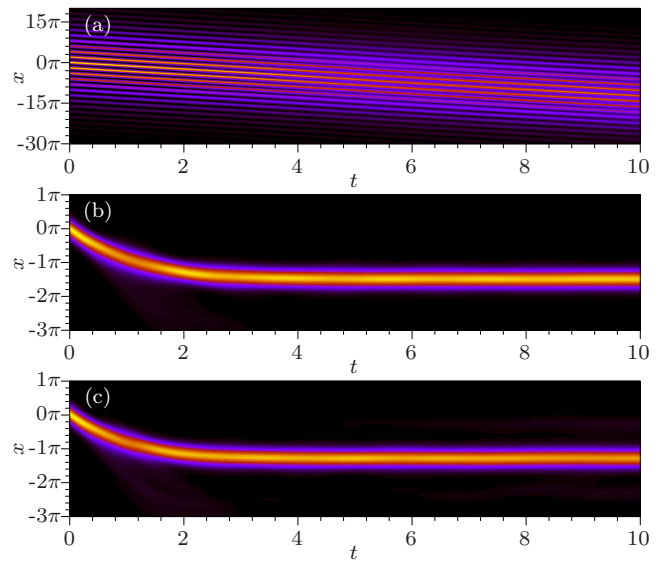}
            \par\end{centering}
        \caption{Mobility of the stable gap solitons. Time evolutions of the atomic
            density $\left|\psi\left(x,t\right)\right|^{2}$ for three different
            solitons with initial speed $v_{0}=4.0$ are plotted in the three
            panels. The three solitons are fundamental soliton near the lower
            edge of the first energy gap (a), deep in the first energy gap (b), and
            deep in the second energy gap (c) respectively. Their wavefunctions,
            effective optical lattice potential, and energy band structures have
            previously been shown in panels (a,b,f) of figure \ref{fig:FundamentalSolitons}.
            \label{fig:Mobility}}
    \end{figure}
    
    \begin{figure}
        \begin{centering}
            \includegraphics{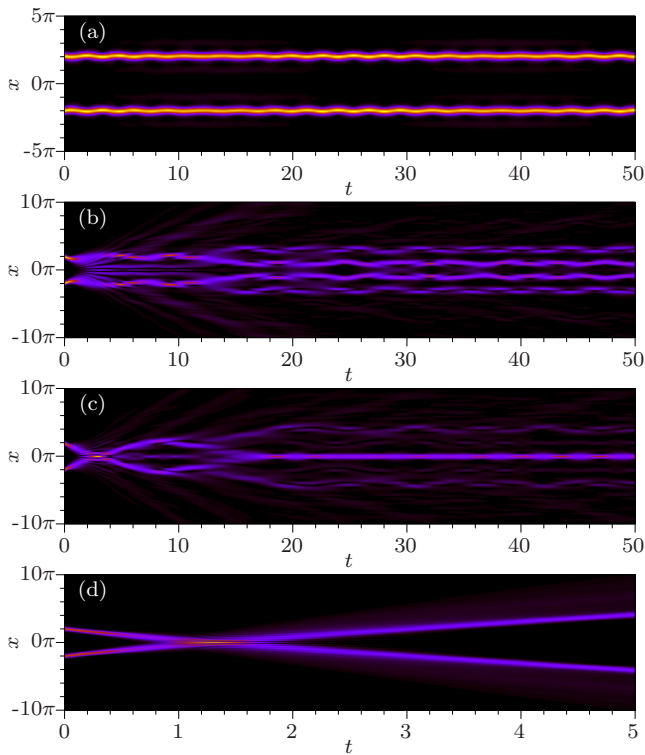}
            \par\end{centering}
        \caption{Collision of two gap solitons (with chemical potential deep in the
            first energy gap). Time evolutions of the atomic density $\left|\psi\left(x,t\right)\right|^{2}$
            are plotted. From top to bottom, the initial colliding velocities
            are set to $v_{0}=0.5$ (a), $2.5$ (b), $3.5$ (c) and $6.0$ (d)
            respectively. \label{fig:Collision}}
    \end{figure}
    
    \subsection{Mobility\label{subsec:Mobility}}
    In this part, we study mobility of the stable gap solitons. In a ring
    cavity, lights of the two counter-propagating modes can have independent
    phases, when their phase difference changes, the optical lattice potential
    produced by their interference will move. Furthermore, the light field
    is built up by pumping the BEC. So, it would be reasonable to expect
    that when the BEC wave packet moves, the optical lattice potential
    will move accordingly, and will put no extra force on the moving BEC,
    such that the BEC can move freely. However, 
    this has been demonstrated to be only partially true. 
    More comprehensive studies show that because the light
    field can not follow the BEC dynamics instantaneously, the optical
    lattice will fall behind the BEC for a certain distance, and will
    put a friction force on the BEC, as a result, the BEC will usually
    undergo a decelerating motion \cite{Qin2020Self,Gietka2019Supersolid}.
    
    To verify the above discussions, we studied the moving dynamics numerically.
    We give an initial velocity $v_{0}$ to the gap soliton by imprinting
    a phase factor $\exp\left[iv_{0}x\right]$ on its wavefunction, then
    examining the afterward time evolution. The results are shown in
    figure \ref{fig:Mobility}. In panels (b,c), for solitons with chemical
    potential deep in both the first (b) and second (c) energy gaps, we
    do observe a decelerating motion of the soliton wave packet. In panel
    (a), for the soliton with chemical potential near the gap edge, the
    deceleration is not obvious, it undergoes an almost free motion. This
    is because in this case the effective optical lattice is very weak
    [see panel (a2) of figure \ref{fig:FundamentalSolitons}], as
    a result, the friction force is also very small, and its deceleration effect
    is hard to be obviously observed on the graph. 
    
    \subsection{Collision\label{subsec:Collision}}
    At last, we show the collision dynamics of two such supersolid gap solitons.
    In figure \ref{fig:Collision}, taking the collisions of two solitons
    with chemical potential deep in the first energy gap as an example, the time
    evolutions of atomic density are plotted for different collision velocities.
    Although a single such gap soliton is movable (as having been shown
    in section \ref{subsec:Mobility}), we found that when the initial
    colliding velocity is small, two solitons can not approach each other,
    they only oscillate around their initial locations with a small amplitude,
    see panel (a). For the medium velocity collision [panels (b,c)],
    the two solitons strongly interact with each other. After some time,
    the two solitons either break into many small pieces [panel (b)],
    or merge into a single wavepacket accompanied by a scattering loss
    of the atoms [panel (c)]. For the large velocity collision [panel
    (d)], the two solitons collide similar to two classical particles,
    however suffering a spatial spreading. Similar collision phenomena have already
    been observed for non-soliton wavepackets in the same model \cite{Qin2021Collision},
    the explanations also apply here.
    
    Strictly speaking, solitons refers to non-spreading localized wavepackets 
    which can interact with other solitons, and emerge from the collision
    unchanged, except for a phase shift \cite{Drazin1989Solitons}. 
    %This is obviously not true here. 
    In this sense, the wavepackets in this
    work would better be called solitary waves. However, in many cases,
    the collision requirement is often given up, and the term soliton may be used
    instead of solitary wave \cite{Akozbek1998Optical}, here we also
    follow such a relaxed definition.
    
    \section{Summary\label{sec:Summary}}
    In summary, we predict that there exist supersolid gap solitons
    in a BEC and optical ring cavity coupling system. We
    studied the system within the mean-field theory, and numerically found
    a few families of gap soliton solutions---fundamental gap solitons
    in both the first and second energy gaps, sub-fundamental gap solitons
    in the second energy gap, and high order gap solitons which consist of
    several fundamental or sub-fundamental solitons. The stability of
    these gap solitons has been checked by numerically simulating the
    time-dependent mean-field equations. The numerical
    results suggest that, for the fundamental solitons, 
    their stability roughly obeys the VK criterion,
    i.e., usually they are stable when their chemical 
    potential $\mu$ is negatively
    dependent on the atom number $N$ ($d\mu/dN<0$), 
    however with some exceptions. All the sub-fundamental
    gap solitons are found to be unstable. The high order gap solitons
    have similar stability as their fundamental or sub-fundamental
    components. For the mobility property, given an initial velocity, these gap
    solitons usually undergo a decelerating motion due to the friction
    force from the light fields 
    (the deceleration may be unobvious when friction force is weak). 
    We also studied the two solitons
    collision dynamics, which are found to be strongly velocity-dependent.
    For small velocity collision, the two solitons can only oscillate
    around their initial location with a small amplitude. For medium velocity
    collision, the two solitons either break into many small pieces,
    or merge into a single wavepacket with a loss of atoms. And the
    large velocity collision behaves similarly to the collision of two classical
    particles, except that the two soliton wavepackets suffer a spatial spreading.
    
    At last, we note that the BEC and optical ring cavity coupling system has already been realized, and supersolid phase in the system has also been identified \cite{Schuster2020Supersolid}. Therefore, the supersolid gap solitons and their dynamical properties reported in this article are ready to be observed experimentally.
    
    \begin{acknowledgments}
        The authors acknowledge supports from the National Natural Science
        Foundation of China (Grants No. 11904063, No. 12074120, No. 11847059,
        and No. 11374003), and the Natural Science Foundation of Shanghai
        (Grant No. 20ZR1418500).
    \end{acknowledgments}

\end{document}